\documentclass[12pt,onecolumn,twoside]{IEEEtran}

\usepackage{amsmath,amssymb,euscript,yfonts,psfrag,latexsym,dsfont,graphicx,bbm,color,amstext,wasysym,pdfsync,graphicx}

\begin{document}

\title{On time-reversibility of linear stochastic models\thanks{This research was supported by grants from AFOSR, NSF, VR, and the SSF.}}

\author{Tryphon T. Georgiou\thanks{Department of Electrical \& Computer Engineering,
University of Minnesota, Minneapolis, Minnesota, {\tt tryphon@umn.edu}}
and Anders Lindquist\thanks{Department of  Automation, Shanghai Jiao Tong University, Shanghai, China, and Center for Industrial and Applied Mathematics and \hspace*{5pt} ACCESS Linnaeus Center,
KTH Royal Institute of Technology, Stockholm, Sweden, {\tt alq@kth.se}}
} 

\maketitle

\begin{abstract}
Reversal of the time direction in stochastic systems driven by white noise has been central throughout the development of stochastic realization theory, filtering and smoothing. Similar ideas were developed in connection with certain problems in the theory of moments, where a duality induced by time reversal was introduced to parametrize solutions. In this latter work it was shown that stochastic systems driven by arbitrary second-order stationary processes can be similarly time-reversed. By combining these two sets of ideas we present herein a generalization of time-reversal in stochastic realization theory. 
\end{abstract}

\newcommand{\mR}{{\mathbb R}}
\newcommand{\mZ}{{\mathbb Z}}
\newcommand{\mN}{{\mathbb N}}
\newcommand{\mE}{{\mathbb E}}
\newcommand{\mC}{{\mathbb C}}
\newcommand{\mD}{{\mathbb D}}
\newcommand{\bU}{{\mathbf U}}
\newcommand{\bW}{{\mathbf W}}
\newcommand{\cF}{{\mathcal F}}

\newcommand{\trace}{{\rm trace}}
\newcommand{\rank}{{\rm rank}}
\newcommand{\Real}{{\Re}e\,}
\newcommand{\half}{{\frac12}}

\section{Introduction}
Time reversal of stochastic systems is central in stochastic realization theory (see, e.g., \cite{Lindquist-P-79,
Lindquist-P-85a,Lindquist-P-85,Lindquist-P-91}, \cite{pavon1980stochastic}, \cite{Lindquist-Pa-84}, \cite{Michaletzky-B-V-98}, \cite{Michaletzky-F-95}),
filtering (see \cite{Lindquist-74}),
smoothing (see \cite{badawi1979mayne,Badawi-L-P-79}, \cite{ferrante2000minimal}) and system identification.
The principal construction is to model a stochastic process as the output of a linear system driven by a noise process which is assumed to be white in discrete time, and orthogonal-increment in continuous time. In studying the dependence between past and future of the process, it is natural to decompose the interface between past and future in a time-symmetric manner.
This gives rise to systems representations of the process running in either time direction, forward or backward in time. 

In a different context (see \cite{georgiou2007caratheodory}) a certain duality between the two time-directions in modeling a stochastic process was introduced in order to characterize solutions to moment problems. In this new setting the noise-process was general (not necessarily white), and the correspondence between the driving inputs to the two time-opposite models was shown to be captured by suitable dual all-pass dynamics.

In the present note, we combine these two sets of ideas to develop a general framework where
two time-opposite stochastic systems model a given stochastic process. We study the relationship between these systems and the corresponding processes. In particular, we recover as a special case certain results of stochastic realization theory
(\cite{Lindquist-P-79}, \cite{pavon1980stochastic}, \cite{badawi1979mayne}) from the 1970's using a novel procedure.

In Section \ref{sec:allpass} we explain how a lifting of state-dynamics into an all-pass system allows direct correspondence between sample-paths of driving generating processes, in opposite time-directions, via causal and anti-causal mappings, respectively. In Section \ref{sec:timereversal} we utilize this mechanism in the context of general output processes and, similarly, introduce a pair of time-opposite models. Finally, in Section \ref{sec:conclusions}, we draw connection to literature on time reversibility and related issues in physics, and we indicate directions for future research.

\section{State dynamics and all-pass extension}\label{sec:allpass}

In this paper we consider discrete-time as well as continuous-time stochastic linear state-dynamics. As usual, in discrete-time these take the form of a set of difference equations
\begin{align}\label{eq:model_discrete}
x(t+1)=Ax(t)+Bu(t)
\end{align}
where $t\in\mZ$, $A\in\mR^{n\times n}, B\in\mR^{n\times p}$,
$n,p\in\mN$, $A$ has all eigenvalues in the open unit disc $\mD=\{z\mid |z|<1\}$, and $u(t),x(t)$ are stationary vector-valued stochastic processes. The system of equations is assumed to be reachable, i.e.,
\begin{align}\label{eq:controllability}
\rank\left[ B,\,AB,\,\ldots A^{n-1}B\right]=n,
\end{align}
and non-trivial in the sense that $B$ is full rank.

In continuous-time, state-dynamics take the form of a system of stochastic differential equations
\begin{align}\label{eq:model_continuous}
dx(t)=Ax(t)dt+Bdu(t)
\end{align}
where, here, $u(t),x(t)$ are stationary continuous-time vector-valued stochastic processes.
Reachability (which in this case, is equivalent to controllability) of the pair $(A,B)$ is also assumed throughout
and the condition for this is identical to the one for discrete-time given above (as is well known). In continuous time, stability of the system of equations is equivalent to $A$ having only eigenvalues with negative real part, and will be assumed throughout along with the condition that $B$ has full rank.

In either case, discrete-time or continuous-time, it is possible to define an output equation so that the overall system is all-pass. This is done next. The assumptions of stationarity and constant parameter matrices is made for simplicity of notation and brevity and can be easily removed.

\subsection{All-pass extension in discrete-time}\label{Section21}

Consider the discrete-time Lyapunov equation
\begin{align}\label{discreteLyapunov}
P= APA^\prime + BB^\prime.
\end{align}
Since $A$ has all eigenvalues inside the unit disc of the complex plane and \eqref{eq:controllability} holds, \eqref{discreteLyapunov} has as solution a matrix $P$ which is positive definite.
The state transformation
\begin{align}\label{eq:statetransform}
\xi= P^{-\half} x,
\end{align}
and
\begin{align}
F=P^{-\half} A P^{\half},\;G=P^{-\half} B,\label{eq:similarity}
\end{align}
brings \eqref{eq:model_discrete} into
\begin{align}\label{eq:model_discrete2}
\xi(t+1)=F\xi(t)+Gu(t).
\end{align}

For this new system, the corresponding Lyapunov equation $X=FXF^\prime+GG^\prime$ has $I_n$ as solution,
where $I_n$ denotes the $(n\times n)$ identity matrix. This fact, namely, that
\begin{align}
I_n=FF^\prime + GG^\prime
\end{align}
implies that
this $[F,G]$ can be embedded as part of an orthogonal matrix
\begin{align}\label{eq:U}
U=\left[\begin{array}{cc} F& G\\ H& J\end{array}\right],
\end{align}
i.e., such that $UU^\prime=U^\prime U= I_{n+p}$. 

Define the transfer function
\begin{align}\label{eq:tfdiscrete}
\bU(z):=H(zI_n-F)^{-1}G+J
\end{align}
corresponding to
\begin{subequations}\label{xisystem}
\begin{align}\label{xisystemfirst}
\xi(t+1)&=F\xi(t)+Gu(t)\\
\bar u(t)&=H\xi(t)+Ju(t).
\end{align}
\end{subequations}
This is also the transfer function of
\begin{subequations}\label{xsystem}
\begin{align}
x(t+1)&=Ax(t)+Bu(t)\\
\bar u(t)&=\bar{B}' x(t)+Ju(t),
\end{align}
\end{subequations}
where $\bar{B}:=P^{-\half}H'$, since the two systems are related by a similarity transformation. Hence,
\begin{align}\label{eq:tfdiscrete2}
\bU(z)=\bar{B}'(zI_n-A)^{-1}B+J.
\end{align}
We claim that $\bU(z)$
is an all-pass transfer function (with respect to the unit disc), i.e.,
that $\bU(z)$ is a transfer function of a stable system (obvious) and that
\begin{align}\label{eq:discreteallpass}
\bU(z)\bU(z^{-1})^\prime=\bU(z^{-1})^\prime \bU(z)=I_p.
\end{align}

The latter claim is immediate after we observe that, since $U^\prime U=I_{n+p}$,
\[
U^\prime\left[\begin{array}{c} \xi(t+1)\\ \bar u(t)\end{array}\right]=\left[\begin{array}{c} \xi(t)\\ u(t)\end{array}\right],
\]
and hence,
\begin{subequations}\label{inversexisystem}
\begin{align}
\xi(t)&=F^\prime\xi(t+1)+H^\prime \bar u(t)\\
u(t)&=G^\prime \xi(t+1)+J^\prime \bar u(t)
\end{align}
\end{subequations}
or, equivalently,
\begin{subequations}\label{inversexsystemprel}
\begin{align}
x(t)&=PA^\prime P^{-1}x(t+1)+P^{\half}H^\prime u(t)\\
u(t)&=B^\prime P^{-1} x(t+1)+J^\prime \bar u(t).
\end{align}
\end{subequations}
Setting 
\begin{equation}
\label{xbar}
\bar{x}(t):=P^{-1} x(t+1),
\end{equation}
\eqref{inversexsystemprel} can be written
\begin{subequations}\label{inversexsystem}
\begin{align}\label{inversexsystema}
\bar{x}(t-1)&=A^\prime\bar{x}(t)+\bar B \bar u(t)\\
u(t)&=B^\prime \bar{x}(t)+J^\prime \bar u(t)\label{inversexsystemb}
\end{align}
\end{subequations}
with transfer function
\begin{align}\label{Ustardefinition}
\bU(z)^*=B^\prime(z^{-1}I_n-A^\prime)^{-1}\bar{B}+J^\prime.
\end{align}
Either of the above systems inverts the dynamical relation $u\to \bar u$ (in \eqref{xsystem} or \eqref{xisystem}).

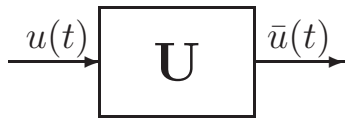
\begin{figure}[htb]
\begin{center}
\setlength{\unitlength}{.008in}
\parbox{304\unitlength}
{\begin{picture}(100,90)
\thicklines
\put(32,60){\makebox(0,0){{\large $u(t)$}}}
\put(0,45){\vector(1,0){60}}
\put(60,10){\framebox(100,70){{\LARGE
\hspace*{7pt}$\bU$ 
}}}
\put(190,60){\makebox(0,0){{\large $\bar u(t)$}}}
\put(160,45){\vector(1,0){60}}
\end{picture}
}
\end{center}
\caption{Realization \eqref{xsystem} in the forward time-direction.}
\label{forwardfigure}
\end{figure}
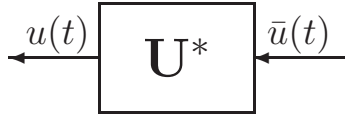
\begin{figure}[htb]
\begin{center}
\setlength{\unitlength}{.008in}
\parbox{304\unitlength}
{\begin{picture}(100,90)
\thicklines
\put(32,60){\makebox(0,0){{\large $u(t)$}}}
\put(60,45){\vector(-1,0){60}}
\put(60,10){\framebox(100,70){{\LARGE
\hspace*{7pt}$\bU^*$ 
}}}
\put(190,60){\makebox(0,0){{\large $\bar u(t)$}}}
\put(220,45){\vector(-1,0){60}}
\end{picture}
}
\end{center}
\caption{Realization \eqref{inversexsystem} in the backward time-direction.}
\label{backwardfigure}
\end{figure}

\subsection{All-pass extension in continuous-time}\label{Section22}

Consider the continuous-time Lyapunov equation
\begin{align}\label{continuousLyapunov}
AP+PA^\prime + BB^\prime =0.
\end{align}
Since $A$ has all its eigenvalues in the left half of the complex plane and since \eqref{eq:controllability} holds, \eqref{continuousLyapunov} has as solution a positive definite matrix $P$.
Once again, applying (\ref{eq:statetransform}-\ref{eq:similarity}),
the system in \eqref{eq:model_continuous}
becomes
\begin{subequations}
\begin{align}\label{eq:model_continuous2}
d\xi(t)=F\xi(t)dt+Gdu(t).
\end{align}
We now seek a completion by adding an output equation
\begin{align}\label{eq:model_output2}
d\bar u(t)=H\xi(t)dt+Jdu(t)
\end{align}
\end{subequations}
so that the transfer function
\begin{align}
\bU(s):=H(sI_n-F)^{-1}G + J
\end{align}
is all-pass
(with respect to the imaginary axis), i.e.,
\begin{align}
\bU(s)\bU(-s)^\prime = \bU(-s)^\prime\bU(s)=I_p.
\end{align}

For this new system, the corresponding Lyapunov equation has as solution the identity matrix and hence,
\begin{align}\label{continuousLyapunov2}
F+F^\prime + GG^\prime =0.
\end{align}
Utilizing this relationship we
note that
\begin{align*}
&(sI_n-F)^{-1}GG^\prime (-sI_n-F^\prime)^{-1}\\
&=(sI_n-F)^{-1}(sI_n-F -sI_n-F^\prime)(-sI_n-F^\prime)^{-1}\\
&=(sI_n-F)^{-1}+(-sI_n-F^\prime)^{-1},
\end{align*}
and we calculate that
\begin{align*}
&\bU(s)\bU(-s)^\prime\\
&= (H (sI_n-F)^{-1}G+J)(G^\prime(-sI_n-F^\prime)^{-1}H^\prime +J^\prime)\\
&=JJ^\prime +H(sI_n-F)^{-1}(GJ^\prime+H^\prime)\\
&\hspace*{1cm}(JG^\prime + H)(-sI_n-F^\prime)^{-1}H^\prime.
\end{align*}
For the product to equal the identity,
\begin{align*}
&JJ^\prime=I_p\\
&H=-JG^\prime.
\end{align*}
Thus, we may take
\begin{align*}
&J=I_p\\
&H=-G^\prime,
\end{align*}
and the forward dynamics
\begin{subequations}\label{eq:continuousforward}
\begin{align}\label{eq:model_continuous2a}
d\xi(t)=F\xi(t)dt+Gdu(t)\phantom{x}\\
\label{eq:model_output2a}
d\bar u(t)=-G^\prime\xi(t)dt+du(t).
\end{align}
\end{subequations}
Substituting $F=-F^\prime-GG^\prime$ from \eqref{continuousLyapunov2} into \eqref{eq:model_continuous2a} we obtain the reverse-time dynamics
\begin{subequations}\label{xibackward}
\begin{align}\label{eq:model_continuous2a_reverse}
d\xi(t)=-F^\prime\xi(t)dt+Gd\bar u(t)\\
\label{eq:model_output2a_reverse}
du(t)=G^\prime\xi(t)dt+d\bar u(t).\phantom{xll}
\end{align}
\end{subequations}
Now defining 
\begin{equation}
\label{xbarcont}
\bar{x}(t):=P^{-1}x(t)
\end{equation}
and using \eqref{eq:statetransform} and \eqref{eq:similarity}, \eqref{xibackward} becomes
\begin{subequations}\label{cbackward}
\begin{align}
d\bar{x}(t)=-A^\prime\bar{x}(t)dt+\bar{B}d\bar u(t)\\
du(t)=B^\prime\bar{x}(t)dt+d\bar u(t),\phantom{xx}\label{ubar2ucont}
\end{align}
\end{subequations}
with transfer function
\begin{align}\label{Ustardefinition_continuous}
\bU(s)^*=B^\prime(sI_n+A^\prime)^{-1}\bar{B}+I_p,
\end{align}
where 
\begin{equation}
\label{Bbarcont}
\bar{B}:=P^{-1}B.
\end{equation}
Furthermore, the forward dynamics \eqref{eq:continuousforward} can be expressed in the form
\begin{subequations}\label{cforward}
\begin{align}
dx(t)=Ax(t)dt+Bdu(t)\\
d\bar u(t)=\bar B^\prime x(t)dt+d u(t)\phantom{b}\label{cforwardb}
\end{align}
\end{subequations}
with transfer function
\begin{align}\label{Udefinition_continuous}
\bU(s)=\bar B^\prime(sI_n-A^\prime)^{-1}B+I_p.
\end{align}

\section{Time-reversal of linear stochastic systems}\label{sec:timereversal}

The development so far allows us to draw a connection between two linear stochastic systems having the same output and driven by a pair of arbitrary, but dual, stationary processes $u(t)$ and $\bar u(t)$, one evolving forward in time and one evolving backward in time. When one of these two processes is white noise (or, orthogonal increment process, in continuous-time),
then so is the other. For this special case we recover
results of \cite{Lindquist-P-79} and \cite{pavon1980stochastic}
in stochastic realization theory.

\subsection{Time-reversal of discrete-time stochastic systems}

Consider a stochastic linear system
\begin{subequations}\label{dsystforward}
\begin{align}
&x(t+1)=Ax(t)+Bu(t) \label{dsystforwarda}\\
&\phantom{xxll}y(t)=Cx(t)+Du(t) \label{dsystforwardb}
\end{align}
\end{subequations}
with an $m$-dimensional output process $y$, and $x,u,A,B$ are defined as in Section \ref{Section21}.
All processes are stationary and the system can be thought as evolving forward in time from the remote past ($t=-\infty$). In particular,
\[
\left(\begin{array}{c}x(t+1)\\ y(t)\end{array}\right) \mbox{ is }\cF_t^{u}\mbox{-measurable}
\]
for all $t\in\mZ$, where
$\cF_t^{u}$ is the $\sigma$-algebra generated by $\{u(s)\mid s\leq t\}$.
Next we construct a stochastic system
\begin{subequations}\label{dsystbackward}
\begin{align}
&\bar{x}(t-1)=A'\bar{x}(t)+\bar{B}\bar{u}(t) \label{dsystbackwarda}\\
&\phantom{xxx}y(t)=\bar{C}\bar{x}(t)+ \bar{D}\bar{u}(t),\label{dsystbackwardb}
\end{align}
\end{subequations}
which evolves backward in time from the remote future ($t=\infty$), and for which
\[
\left(\begin{array}{c}\bar x(t-1)\\ y(t)\end{array}\right) \mbox{ is }\bar\cF_t^{\bar u}\mbox{-measurable}
\]
for all $t\in\mZ$, where
$\bar\cF_t^{\bar u}$ is the $\sigma$-algebra generated by $\{\bar u(s)\mid s\geq t\}$. The processes $\bar x,x,\bar u,u$ relate as in the previous section. More specifically, as shown in Section \ref{Section21},
\[
\bar u(t)\mbox{ is }\cF_t^{u}\mbox{-measurable}
\]
while
\[
u(t)\mbox{ is }\bar\cF_t^{\bar u}\mbox{-measurable}
\]
for all $t$, as examplified in Figures \ref{forwardfigure} and \ref{backwardfigure}.

In fact, the all-pass extension \eqref{xsystem} of \eqref{dsystforwarda} yields 
\begin{equation}
\label{w2wbar}
\bar{u}(t)  =  \bar{B}'x(t)+Ju(t)
\end{equation}
It follows from \eqref{inversexsystemb} that \eqref{w2wbar} can be inverted to yield
\begin{equation}
\label{wbar2w}
u(t)=B'\bar{x}(t)+J'\bar{u}(t), 
\end{equation}
where $\bar{x}(t)=P^{-1}x(t+1)$, and that we have the reverse-time recursion
\begin{subequations}\label{backwardagain}
\begin{equation}
\label{xtminus1}
\bar{x}(t-1)=A'\bar{x}(t)+\bar{B}\bar{u}(t).
\end{equation}
Then inserting \eqref{wbar2w} and 
\begin{displaymath}
x(t)=P\bar{x}(t-1)=PA'\bar{x}(t)+P\bar{B}\bar{u}(t)
\end{displaymath}
into \eqref{dsystforwardb}, we obtain 
\begin{equation}
y(t)=\bar{C}\bar{x}(t)+ \bar{D}\bar{u}(t),
\end{equation}
\end{subequations}
where $\bar{D}:=DJ'$ and
\begin{equation}
\label{Cbar}
\bar{C}:= CPA' +DB' .
\end{equation}
Then, \eqref{backwardagain} is precisely what we wanted to establish.

Moreover, the transfer functions
\begin{align}\bW(z)=C(zI_n-A)^{-1}B+D
\end{align}
of \eqref{dsystforward}
and
\begin{align}
\bar\bW(z)=\bar C(z^{-1}I_n-A')^{-1}\bar B+\bar D
\end{align}
of \eqref{dsystbackward} satisfy
\begin{align}
\bW(z)=\bar\bW(z)\bU(z).
\end{align}
In the context of stochastic realization theory, discussed next, $\bU(z)$ is called {\em structural function} (\cite{Lindquist-P-85,Lindquist-P-91}).

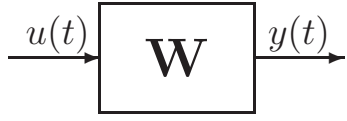
\begin{figure}[h]
\begin{center}
\setlength{\unitlength}{.008in}
\parbox{304\unitlength}
{\begin{picture}(100,90)
\thicklines
\put(32,60){\makebox(0,0){{\large $u(t)$}}}
\put(0,45){\vector(1,0){60}}
\put(60,10){\framebox(100,70){{\LARGE
\hspace*{7pt}$\bW$ 
}}}
\put(190,60){\makebox(0,0){{\large $y(t)$}}}
\put(160,45){\vector(1,0){60}}
\end{picture}
}
\end{center}
\caption{The forward stochastic system \eqref{dsystforward}.}
\label{forwardfigure2}
\end{figure}
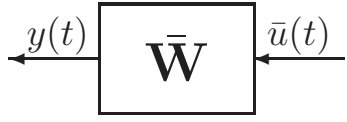
\begin{figure}[h]
\begin{center}
\setlength{\unitlength}{.008in}
\parbox{304\unitlength}
{\begin{picture}(100,90)
\thicklines
\put(32,60){\makebox(0,0){{\large $y(t)$}}}
\put(60,45){\vector(-1,0){60}}
\put(60,10){\framebox(100,70){{\LARGE
\hspace*{7pt}$\bar\bW$ 
}}}
\put(190,60){\makebox(0,0){{\large $\bar u(t)$}}}
\put(220,45){\vector(-1,0){60}}
\end{picture}
}
\end{center}
\caption{The backward stochastic system \eqref{dsystbackward}}
\label{backwardfigure2}
\end{figure}

\subsubsection{Time-reversal of stochastic realizations.}
Given an $m$-dimensional stationary process $y$, consider a  minimal stochastic realization
\eqref{dsystforward},
evolving forward in time, where now $u$ is a normalized white noise process, i.e.,
\[
\mE\{u(t)u(s)'\}=I_p\delta_{t-s}.
\]
Since $\mathbf{U}$, given by \eqref{eq:tfdiscrete2}, is all-pass, $\bar{u}$ is also a normalized white noise process, i.e.,
\[
\mE\{\bar{u}(t)\bar{u}(s)'\}=I_p\delta_{t-s}.
\]
From the reverse-time recursion \eqref{dsystbackwarda}
\[
\bar x(t)=\sum_{k=t+1}^\infty (A')^{k-(t+1)}\bar B \bar u(k).
\]
Since, $\bar u$ is a white noise process, $\mE\{\bar x(t)\bar u(s)'\}=0$ for all $s\leq t$.
Consequently, \eqref{dsystbackward} is a backward stochastic realization in the sense of stochastic realization theory.

\subsection{Time-reversal of continuous-time stochastic systems}

We now turn to the continuous-time case. Let 
\begin{subequations}\label{csystforward}
\begin{align}
&dx=Axdt+Bdu \label{csystforwarda}\\
&dy=Cxdt+Ddu  \label{csystforwardb}
\end{align}
\end{subequations}
be a stochastic system with $x,u,A,B$ as in Section \ref{Section22}, evolving forward in time from the remote past ($t=-\infty$). All processes have stationary increments
and
\[
\left(\begin{array}{c}x(t)\\ y(t)\end{array}\right) \mbox{ is }\cF_t^{u}\mbox{-measurable}
\]
for all $t\in\mR$, where
$\cF_t^{u}$ is the $\sigma$-algebra generated by $\{u(s)\mid s\leq t\}$.

The all-pass extension of Section \ref{Section22} yields
\begin{equation}
d\bar{u}=du -\bar{B}'xdt
\end{equation}
as well as the reverse-time relation
\begin{subequations}\label{cbackward2}
\begin{align}
d\bar{x}=-A^\prime\bar{x}dt+\bar{B}d\bar u\\
du=B^\prime\bar{x}dt+d\bar u,\phantom{xx}\label{ubar2ucont2}
\end{align}
\end{subequations}
where $\bar{x}(t)=P^{-1}x(t)$. Inserting \eqref{ubar2ucont2} into 
\begin{displaymath}
dy=CP\bar{x}dt+Ddu
\end{displaymath}
yields
\begin{displaymath}
dy=\bar{C}\bar{x}dt+Dd\bar{u},
\end{displaymath}
where
\begin{equation}
\bar{C}=CP+DB'.
\end{equation}
Thus, the reverse-time system is
\begin{subequations}\label{csystbackward}
\begin{align}
&d\bar{x}=-A'\bar{x}dt+\bar{B}d\bar{u} \label{csystbackwarda}\\
&dy=\bar{C}\bar{x}dt+Dd\bar{u}. \label{csystbackwardb}
\end{align}
\end{subequations}
From this, we deduce that
\[
\left(\begin{array}{c}\bar x(t)\\ y(t)\end{array}\right) \mbox{ is }\bar\cF_t^{\bar u}\mbox{-measurable}
\]
for all $t\in\mR$. We also note that the transfer function
\[
\bW(s)=C(sI_n-A)^{-1}B+D
\]
of \eqref{csystforward} and the transfer function
\[
\bar\bW(s)=\bar C(sI_n+A')^{-1}\bar B+D
\]
of \eqref{csystbackward} also satisfy
\[
\bW(s)=\bar\bW(s)\bU(s)
\]
as in discrete-time.

\subsubsection{Time-reversal of stochastic realizations.}
In continuous-time stochastic realization theory, \eqref{csystforward} is a forward minimal stochastic realization of an $m$-dimensional process $y$ with stationary increments provided $u$ is a normalized orthogonal-increment process
satisfying
\[ 
 \mE\{ du(t)du(t)'\}=I_pdt.
\]
Since $\bU(s)$ is all-pass,  
\begin{equation}
d\bar{u}=du -\bar{B}'xdt
\end{equation}
also defines a stationary orthogonal-increment process $\bar u$ such that 
\[
\mE\{ d\bar{u}(t)d\bar{u}(t)'\}=I_pdt. 
\]
It remains to show that \eqref{csystbackward}
is a backward stochastic realization, that is, at each time $t$ the past increments of $\bar u$ are orthogonal to $\bar x(t)$.
But this follows from the fact that
\[
\bar x(t)=\int_t^\infty e^{-A' (t-s)}\bar Bd\bar u(s)
\]
and $\bar u$ has orthogonal increments.

\section{Concluding remarks}\label{sec:conclusions}
The direction of time in physical laws and the fact that physical laws are symmetric with respect to time have occupied some of the most prominent minds in science and mathematics (\cite{schrodinger1931umkehrung}, \cite{kolmogorov1992selected}, \cite{shiryayev1992reversibility}). These early consideration were motivated by no less an issue than that of the very nature of the quantum. Indeed, Erwin Schr\"odinger's aim appears to have been to draw a classical analog to his famous equation.
A large body of work followed.

In particular, closer to our immediate interests, dual time-reversed models have been employed to model, in different time-directions, Brownian or Schr\"odinger bridges (see \cite{pavon1991free}, \cite{dai1990markov}), a subject which is related to reciprocal processes (\cite{jamison1974reciprocal}, \cite{Krener-86}, \cite{levy1990modeling}, \cite{dai1991stochastic}).
The topic of time reversibility has also been central to thermodynamics,
and in recent years studies have sought to elucidate its relation to systems theory (see \cite{haddad2008time,haddad2009thermodynamics}).  Possible connections between this body of work and our present paper will be the subject of future work.

The thesis of the present work is that under mild assumptions on a stochastic process,
any model that consists of a linear stable dynamical system driven by an appropriate input process can be reversed in time. In fact, a reverse-time dual system along with the corresponding input process can be obtained via an all-pass extension of the state equation. The correspondence between the two input processes can be expressed
in terms of each other by a causal and an anti-causal map, respectively.

The formalism of our paper can easily be extended to a non-stationary setting at a price of increased notational, but not conceptual, complexity. Informally, and in order to underscore the point, if $u(t)$ is a non-stationary process and the linear system is time-varying, under suitable conditions, a reverse-time system and a process $\bar u(t)$ can be similarly constructed
via a time-varying orthogonal transformation.

\bibliographystyle{IEEEtran}
\bibliography{referencesfile}                                                        

\begin{thebibliography}{10}
\providecommand{\url}[1]{#1}
\csname url@samestyle\endcsname
\providecommand{\newblock}{\relax}
\providecommand{\bibinfo}[2]{#2}
\providecommand{\BIBentrySTDinterwordspacing}{\spaceskip=0pt\relax}
\providecommand{\BIBentryALTinterwordstretchfactor}{4}
\providecommand{\BIBentryALTinterwordspacing}{\spaceskip=\fontdimen2\font plus
\BIBentryALTinterwordstretchfactor\fontdimen3\font minus
  \fontdimen4\font\relax}
\providecommand{\BIBforeignlanguage}[2]{{%
\expandafter\ifx\csname l@#1\endcsname\relax
\typeout{** WARNING: IEEEtran.bst: No hyphenation pattern has been}%
\typeout{** loaded for the language `#1'. Using the pattern for}%
\typeout{** the default language instead.}%
\else
\language=\csname l@#1\endcsname
\fi
#2}}
\providecommand{\BIBdecl}{\relax}
\BIBdecl

\bibitem{Lindquist-P-79}
A.~Lindquist and G.~Picci, ``On the stochastic realization problem,''
  \emph{SIAM J. Control Optim.}, vol.~17, no.~3, pp. 365--389, 1979.

\bibitem{Lindquist-P-85a}
------, ``Forward and backward semimartingale models for {G}aussian processes
  with stationary increments,'' \emph{Stochastics}, vol.~15, no.~1, pp. 1--50,
  1985.

\bibitem{Lindquist-P-85}
------, ``Realization theory for multivariate stationary {G}aussian
  processes,'' \emph{SIAM J. Control Optim.}, vol.~23, no.~6, pp. 809--857,
  1985.

\bibitem{Lindquist-P-91}
------, ``A geometric approach to modelling and estimation of linear stochastic
  systems,'' \emph{J. Math. Systems Estim. Control}, vol.~1, no.~3, pp.
  241--333, 1991.

\bibitem{pavon1980stochastic}
M.~Pavon, ``Stochastic realization and invariant directions of the matrix
  {R}iccati equation,'' \emph{SIAM Journal on Control and Optimization},
  vol.~18, no.~2, pp. 155--180, 1980.

\bibitem{Lindquist-Pa-84}
A.~Lindquist and M.~Pavon, ``On the structure of state-space models for
  discrete-time stochastic vector processes,'' \emph{IEEE Trans. Automat.
  Control}, vol.~29, no.~5, pp. 418--432, 1984.

\bibitem{Michaletzky-B-V-98}
G.~Michaletzky, J.~Bokor, and P.~V{\'a}rlaki, \emph{Representability of
  stochastic systems}.\hskip 1em plus 0.5em minus 0.4em\relax Budapest:
  Akad\'emiai Kiad\'o, 1998.

\bibitem{Michaletzky-F-95}
G.~Michaletzky and A.~Ferrante, ``Splitting subspaces and acausal spectral
  factors,'' \emph{J. Math. Systems Estim. Control}, vol.~5, no.~3, pp. 1--26,
  1995.

\bibitem{Lindquist-74}
A.~Lindquist, ``A new algorithm for optimal filtering of discrete-time
  stationary processes,'' \emph{SIAM J. Control}, vol.~12, pp. 736--746, 1974.

\bibitem{badawi1979mayne}
F.~Badawi, A.~Lindquist, and M.~Pavon, ``On the {M}ayne-{F}raser smoothing
  formula and stochastic realization theory for nonstationary linear stochastic
  systems,'' in \emph{Decision and Control including the Symposium on Adaptive
  Processes, 1979 18th IEEE Conference on}, vol.~18.\hskip 1em plus 0.5em minus
  0.4em\relax IEEE, 1979, pp. 505--510.

\bibitem{Badawi-L-P-79}
F.~A. Badawi, A.~Lindquist, and M.~Pavon, ``A stochastic realization approach
  to the smoothing problem,'' \emph{IEEE Trans. Automat. Control}, vol.~24,
  no.~6, pp. 878--888, 1979.

\bibitem{ferrante2000minimal}
A.~Ferrante and G.~Picci, ``Minimal realization and dynamic properties of
  optimal smoothers,'' \emph{Automatic Control, IEEE Transactions on}, vol.~45,
  no.~11, pp. 2028--2046, 2000.

\bibitem{georgiou2007caratheodory}
T.~T. Georgiou, ``The {C}arath{\'e}odory--{F}ej{\'e}r--{P}isarenko
  decomposition and its multivariable counterpart,'' \emph{Automatic Control,
  IEEE Transactions on}, vol.~52, no.~2, pp. 212--228, 2007.

\bibitem{schrodinger1931umkehrung}
E.~Schr{\"o}dinger, \emph{{\"U}ber die Umkehrung der Naturgesetze}.\hskip 1em
  plus 0.5em minus 0.4em\relax Akad. d. Wissenschaften, 1931.

\bibitem{kolmogorov1992selected}
A.~Kolmogorov, \emph{Selected Works of AN Kolmogorov: Probability theory and
  mathematical statistics}.\hskip 1em plus 0.5em minus 0.4em\relax Springer,
  1992, vol.~26.

\bibitem{shiryayev1992reversibility}
A.~Shiryayev, ``On the reversibility of the statistical laws of nature,'' in
  \emph{Selected Works of AN Kolmogorov}.\hskip 1em plus 0.5em minus
  0.4em\relax Springer, 1992, pp. 209--215.

\bibitem{pavon1991free}
M.~Pavon and A.~Wakolbinger, ``On free energy, stochastic control, and
  {S}chr{\"o}dinger processes,'' in \emph{Modeling, Estimation and Control of
  Systems with Uncertainty}.\hskip 1em plus 0.5em minus 0.4em\relax Springer,
  1991, pp. 334--348.

\bibitem{dai1990markov}
P.~Dai~Pra and M.~Pavon, ``On the {M}arkov processes of {S}chr{\"o}dinger, the
  {F}eynman-{K}ac formula and stochastic control,'' in \emph{Realization and
  Modelling in System Theory}.\hskip 1em plus 0.5em minus 0.4em\relax Springer,
  1990, pp. 497--504.

\bibitem{jamison1974reciprocal}
B.~Jamison, ``Reciprocal processes,'' \emph{Probability Theory and Related
  Fields}, vol.~30, no.~1, pp. 65--86, 1974.

\bibitem{Krener-86}
A.~Krener, ``Reciprocal processes and the stochastic realization problem for
  acausal systems,'' in \emph{Modelling, Identification and Robust Control},
  C.~I. Byrnes and A.~Lindquist, Eds.\hskip 1em plus 0.5em minus 0.4em\relax
  Amsterdam: North-Holland, 1986, pp. 197--211.

\bibitem{levy1990modeling}
B.~C. Levy, R.~Frezza, and A.~J. Krener, ``Modeling and estimation of
  discrete-time gaussian reciprocal processes,'' \emph{Automatic Control, IEEE
  Transactions on}, vol.~35, no.~9, pp. 1013--1023, 1990.

\bibitem{dai1991stochastic}
P.~Dai~Pra, ``A stochastic control approach to reciprocal diffusion
  processes,'' \emph{Applied mathematics and Optimization}, vol.~23, no.~1, pp.
  313--329, 1991.

\bibitem{haddad2008time}
W.~M. Haddad, V.~Chellaboina, and S.~G. Nersesov, ``Time-reversal symmetry,
  poincar{\'e} recurrence, irreversibility, and the entropic arrow of time:
  From mechanics to system thermodynamics,'' \emph{Nonlinear Analysis: Real
  World Applications}, vol.~9, no.~2, pp. 250--271, 2008.

\bibitem{haddad2009thermodynamics}
------, \emph{Thermodynamics: A dynamical systems approach}.\hskip 1em plus
  0.5em minus 0.4em\relax Princeton University Press, 2009.

\end{thebibliography}
\end{document}